\documentclass[preprint,preprintnumbers,amsmath,amssymb]{revtex4}
\usepackage{amssymb,latexsym,amsmath}
\usepackage[activeacute,english]{babel}
\usepackage{fancyhdr}
\usepackage{textcomp}
\usepackage{amsfonts}
\usepackage{graphicx}
\usepackage{subfigure}
\usepackage{graphics}
\usepackage{color}
\usepackage{array}

\begin{document}

\title{Three-loop Correction to the Instanton Density. II. The Sine-Gordon potential}

\author{M.A.~Escobar-Ruiz$^1$}
\email{mauricio.escobar@nucleares.unam.mx}

\author{E.~Shuryak$^2$}
\email{edward.shuryak@stonybrook.edu}

\author{A.V.~Turbiner$^{1,2}$}
\email{turbiner@nucleares.unam.mx, alexander.turbiner@stonybrook.edu}

\affiliation{$^1$ Instituto de Ciencias Nucleares, Universidad Nacional Aut\'onoma de M\'exico,
Apartado Postal 70-543, 04510 M\'exico, D.F., M\'exico}

\affiliation{$^2$  Department of Physics and Astronomy, Stony Brook University,
Stony Brook, NY 11794-3800, USA}

\date{November 1, 2015}

\begin{abstract}
In this second paper on quantum fluctuations near the classical instanton configurations, see \cite{DWP},
we focus on another well studied quantum-mechanical problem, the one-dimensional Sine-Gordon potential
(the Mathieu potential).
Using only the tools from quantum field theory, the Feynman diagrams in the instanton background,
we calculate the tunneling amplitude (the instanton density) to the three-loop order. The result
confirms (to seven significant figures) the one recently recalculated by G. V. Dunne and M.
\"{U}nsal, {\it Phys. Rev. \bf D 89}, 105009 (2014) from the resurgence perspective.
As in the double well potential case, we found that the largest contribution is given by the diagrams originating from the Jacobian. We again observe that in the three-loop case individual Feynman diagrams contain irrational contributions,
while their sum does not.
\end{abstract}

\maketitle

\section*{Introduction}

\hspace{0.4cm}
Since it is our second paper of the series, following the one on the double-well potential \cite{DWP},
it does not need an extensive introduction. Topological solitons, instantons in particular, are widely used in the context of quantum field theories and condense matter physics.
Their relation to standard perturbative series is an old issue, which continues to produce interesting results, so far mostly in quantum mechanical context.

\hspace{0.4cm}
The Sine-Gordon (SG) field theory has been extensively studied in classical context, with an enormous literature dedicated to it, see e.g. \cite{Scott} and references therein. Coleman \cite{ColemanI} has extended the results to the quantized theory by relating the SG field theory to the zero-charge sector
of the massive Thirring model. Also in \cite{Liang} the explicit calculations for the tunneling amplitude using the so-called nonvacuum instantons at finite energy were presented.

\hspace{0.4cm} The quantum mechanical SG potential (the Mathieu potential) is the basic element of condense matter theory. Tunneling from one minimum to the next, in the path integral formulation, is described by
Euclidean classical paths -- the instantons. The issues we discuss in this paper deal with quantum fluctuations around these paths. We would like to demonstrate by an explicit calculation how our tools work in this -- well controlled and studied setting --before applying them to  more complicated/realistic settings in quantum field theory. Therefore we do $not$ use anything stemming from the Schr\"odinger equation in this work, in particularly do not use series resulting from recurrence relations or resurgence relations
(in general, conjectured) by several authors.

\hspace{0.4cm} One reason to study SG is to explore further the existing deep connections between
the quantum mechanical instantons -- via Schr\"odinger equation -- with wider mathematical issues,
of approximate solutions to differential equations, defined in terms of certain generalized series.
A particular form of an exact quantization condition was \textit{conjectured}
by J.~Zinn-Justin \cite{J. Zinn-Justin}, which links series around the instanton,
instanton/anti-instanton sectors with the usual perturbative series in the perturbative vacuum.
It remains unknown whether it can or cannot be generalized  to the field theory cases we are mainly interested in. Recently, for the quartic double well and Sine-Gordon potentials Dunne and \"{U}nsal
(see \cite{V. Dunne} and also references therein) have presented more arguments for this connection
and made it more precise, which they call {\it resurgent relation} between perturbative and instanton sectors.

\hspace{0.4cm} Another reason for which we decided to do this work is a certain set of observations
about Feynman diagrams on top of the instanton for the double well potential with degenerate minima
we observed in our first paper \cite{DWP}. We wanted to see how general they are, using another example, now
with infinitely many degenerate minima.
The SG potential also has new vertices and thus many new diagrams. As we will show below, indeed all these
trends repeat themselves in this second setting as well.

\hspace{0.4cm} Few comments on the history of present approach.
Omitting well known classic papers on instanton calculus we mention a pioneering paper
\cite{Stone}, where  the two-loop correction to the tunneling amplitude for the SG was calculated. In particular, the formalism for treating the zero-mode singularities was described in detail.

\section*{Three-loop correction to the instanton density}

Let us consider the quantum-mechanical problem of a particle of mass $m=1$ in the Sine-Gordon potential
\begin{equation}
V \ = \   \frac{1}{g^2}[\,  1- \cos(g\,x) \,  ]      \ .
\end{equation}
The well-known instanton solution $X_{inst}(t)\, = \,\frac{4}{g} \arctan(e^{t})$ describes the tunneling between adjacent minima by the Euclidean classical path with the action $S_0=S[X_{inst}(t)]=\frac{8}{g^2}$.
Our notation for the coupling is related to those used in \cite{V. Dunne} by $g_{DU}=\frac{g}{2}$. The classical action $S_0$ of the instanton solution is therefore large and $\frac{1}{S_0}$ is used
in the expansion.

The SG potential has an infinite number of degenerate minima, and perturbative levels in them  form a continuous band, with states within the band labeled by Bloch angular parameter $\theta$. The energy of the lowest band is

\begin{equation}
E_{\theta}^{\text{(lowest \ band)}} \ = \  E_0 - \frac{\delta\,E}{2} \,\cos \theta \ ,
\label{E0band}
\end{equation}

where $E_0$ is the naive ground state energy, without tunneling, written as the following expansion
\begin{equation}
E_0 \ = \ \frac{1}{2}\,\sum_{n=0}^\infty \frac{A_n}{S_0^n} \quad , \qquad (A_0=1) \ ,
\label{E0}
\end{equation}
 while   $ \delta\,E \ = \ E_{\theta=\pi}^{\text{(lowest \ band)}} - E_{\theta=0}^{\text{(lowest \ band)}}$
 generates another series,
related to the so called instanton density
\begin{equation}
\delta\,E \ = \ \Delta E\, \sum_{n=0}^\infty \frac{B_n}{S_0^n}  \quad , \quad (B_0=1) \ ,
\label{delE0}
\end{equation}
here $\Delta E = 2 \sqrt{\frac{2\,S_0}{\pi}}\,e^{-S_0}$ is the well-known one-loop semiclassical
result \cite{Stone}. Coefficients $A_{n}$ in the series (\ref{E0}) can be calculated using the ordinary
perturbation theory (see \cite{Bender}) while many coefficients $B_n$ in the expansion (\ref{delE0})
were found by J.~Zinn-Justin and collaborators, 1981-2005 (see \cite{J. Zinn-Justin} and references therein), obtained via the so called \textit{exact Bohr-Sommerfeld quantization condition}. It was remarkably
refined by Dunne-\"{U}nsal \cite{V. Dunne} in the formalism of the so-called
resurgent trans-series.

Alternatively, using the Feynman diagrams technique Lowe and Stone  \cite{Stone} calculated
the two-loop correction $B_1=-7/8$ which was later on reproduced in \cite{J. Zinn-Justin}
in the so-called {\it exact Bohr-Sommerfeld quantization} technique.
Higher order coefficients $B_n$ in (\ref{delE0}) can also be computed in this way.
Since we calculate the energy difference, all Feynman diagrams in the instanton background
(with the instanton-based vertices and the Green's function) need to be accompanied by subtraction of
the same diagrams for the trivial $x = 0$ saddle point (see \cite{1-E. Shuryak} for details).
For $\frac{1}{\Delta E}\gg \tau \gg 1$ it permits to evaluate the ratio
\[
\frac{\quad \  \langle \, \pi \,| e^{-H\,\tau} | 0 \,\rangle_{x=X_{inst}} }{\langle\,  0 \, | e^{-H\,\tau} | 0\, \rangle_{x=0}}
\]
where the matrix elements
$\langle  \pi  | e^{-H\,\tau} |  0 \rangle_{x=X_{inst}}  \ ,\ \langle  0| e^{-H\,\tau} |  0 \rangle_{x=0}$
are calculated using the instanton-based Green's function and the Green function of the harmonic oscillator,
respectively.

The instanton-based Green's function $G(x,y)$  form to be used
\begin{equation}
\begin{aligned}
G(x,y) \ & = \ \frac{G^0(x,y)}{2\,(1+x^2)(1+y^2)}\bigg[ 1+  4\,x\,y  +x^2\, y^2 + x^2+y^2
\\ &  \  + (  1-4xy+x^2 y^2+ x^2+y^2 +2(1-xy)|x-y|  )\log(2\,G^0(x,y))        \bigg] \ ,
\label{GF}
\end{aligned}
\end{equation}
is expressed in variables $x \,=\, \tanh(\frac{t_1}{2}),\,y \,=\,\tanh(\frac{t_2}{2})\ $,
in which the  familiar Green function $\frac{1}{2}e^{-|t_1-t_2|}$ of the harmonic oscillator is
\begin{equation}
G^0(x,y) \ = \   \frac{1-|x-y|-x\,y}{2(1+|x-y|-x\,y)} \ ,
\label{GF0}
\end{equation}
In its derivation there were two steps. One was  to find a function which satisfies
the Green function equation, used via two linearly-independent solutions and standard Wronskian method.
The second step is related to a zero mode: one can add a term $\phi_0(t_1)\phi_0(t_2)$
with any coefficient and still satisfy the equation. The coefficient is then fixed
from orthogonality to the zero mode.

\begin{figure}[htp]
\begin{center}
\includegraphics[width=4.5in,angle=0]{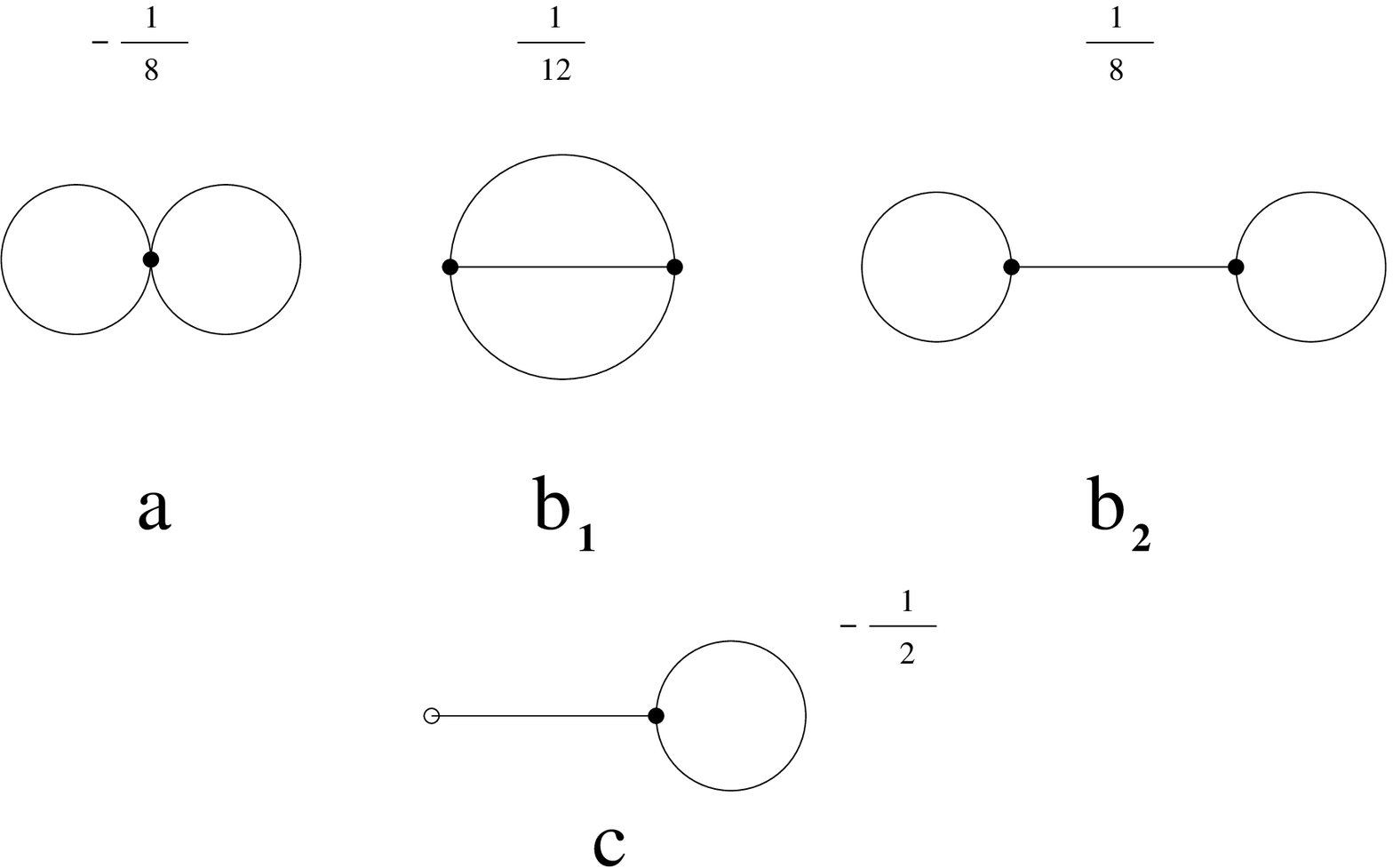}
\caption{Diagrams contributing to the two-loop correction $B_1=a+b_1+b_2+c$. They enter into the coefficient $B_2$ via the term $B_{2loop}$. For the instanton field the effective triple, quartic, quintic and sextic vertices
are $V_3 = 4\,\sqrt{2}\,\frac{\sinh(t)}{\cosh^2(t)}\,S_0^{-1/2}$,
$V_4 = 8\,(\frac{2}{\cosh^{2}t}-1)\,S_0^{-1}$\,,
$V_5 = -32\,\sqrt{2}\,\frac{\sinh(t)}{\cosh^2(t)}\,S_0^{-3/2}$\,,
$V_6 = -64\,(\frac{2}{\cosh^{2}t}-1)\,S_0^{-2}$\ , respectively, all marked by (filled) bullets, while for the subtracted vacuum field diagrams we have $V_3=V_5=0$, $V_4=8\,S_0^{-1}$ and $V_6=64\,S_0^{-2}$\,. The tadpole in diagram $c$, which comes from the zero-mode Jacobian rather than from the action, is effectively represented by the vertex (Jacobian source) $V_{tad} = \frac{1}{\sqrt{2}}\frac{\sinh(t)}{ \cosh^2(t)}\,S_0^{-1/2}$,
marked (unfilled) open bullet (see text). The signs of contributions
and symmetry factors are indicated.}
\label{F1}
\end{center}
\end{figure}

The two-loop coefficient  in (\ref{delE0}) is \cite{Stone}
\[
B_1   =   a+b_1+b_2+c \ ,
\]
\begin{equation}
\begin{aligned}
\label{B1}
  a=-\frac{53}{60}, \qquad b_1=\frac{3}{40}\ ,\qquad b_2=\frac{7}{20}\ ,
  \qquad c=-\frac{5}{12}\ .
\end{aligned}
\end{equation}
\bigskip
reflecting the contribution of four Feynman diagrams, see Fig. \ref{F1}.

\begin{figure}[htp]
\begin{center}
\includegraphics[width=3.5in,angle=0]{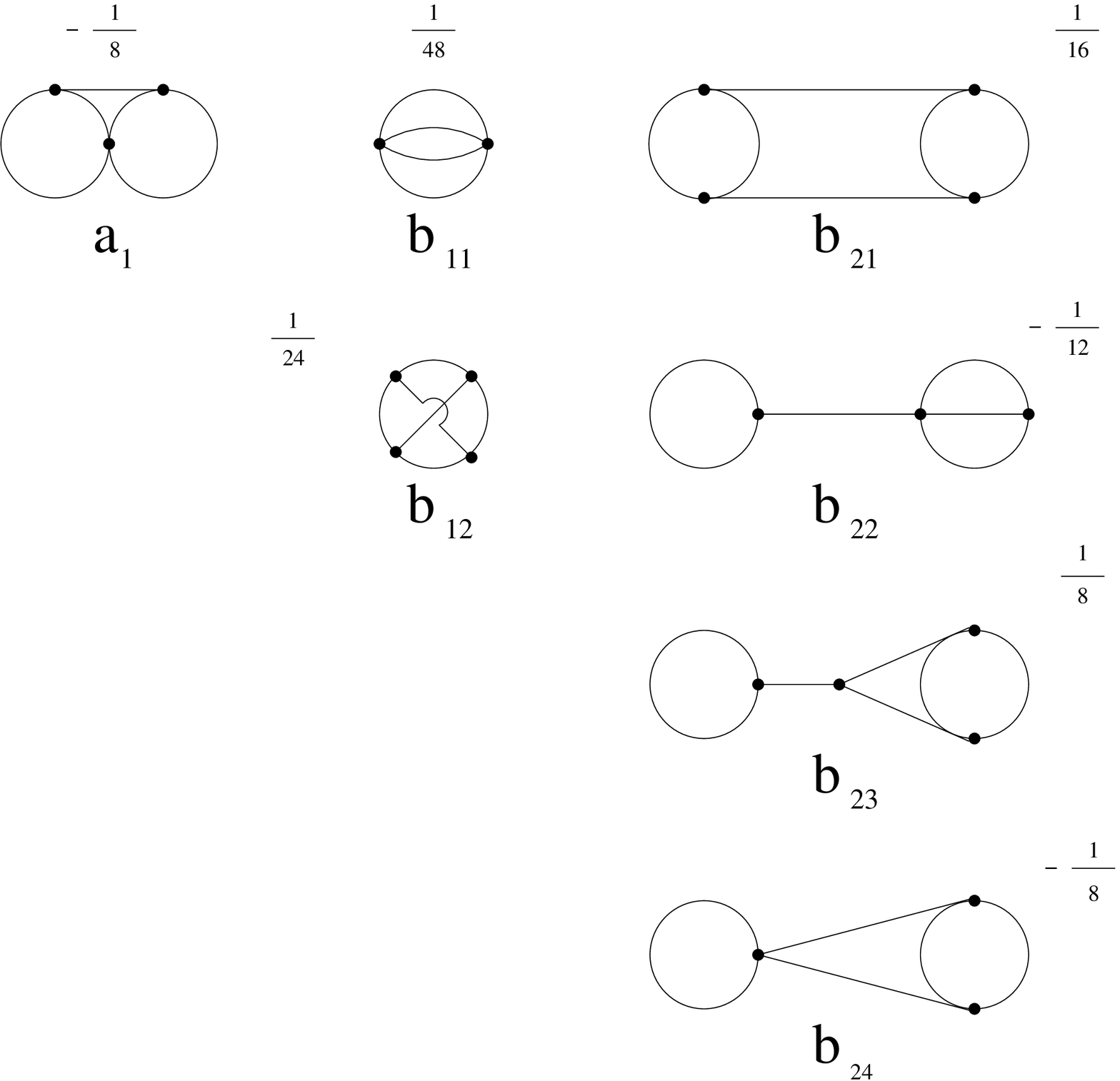}  \vspace{0.8cm}  \\
\includegraphics[width=4.5in,angle=0]{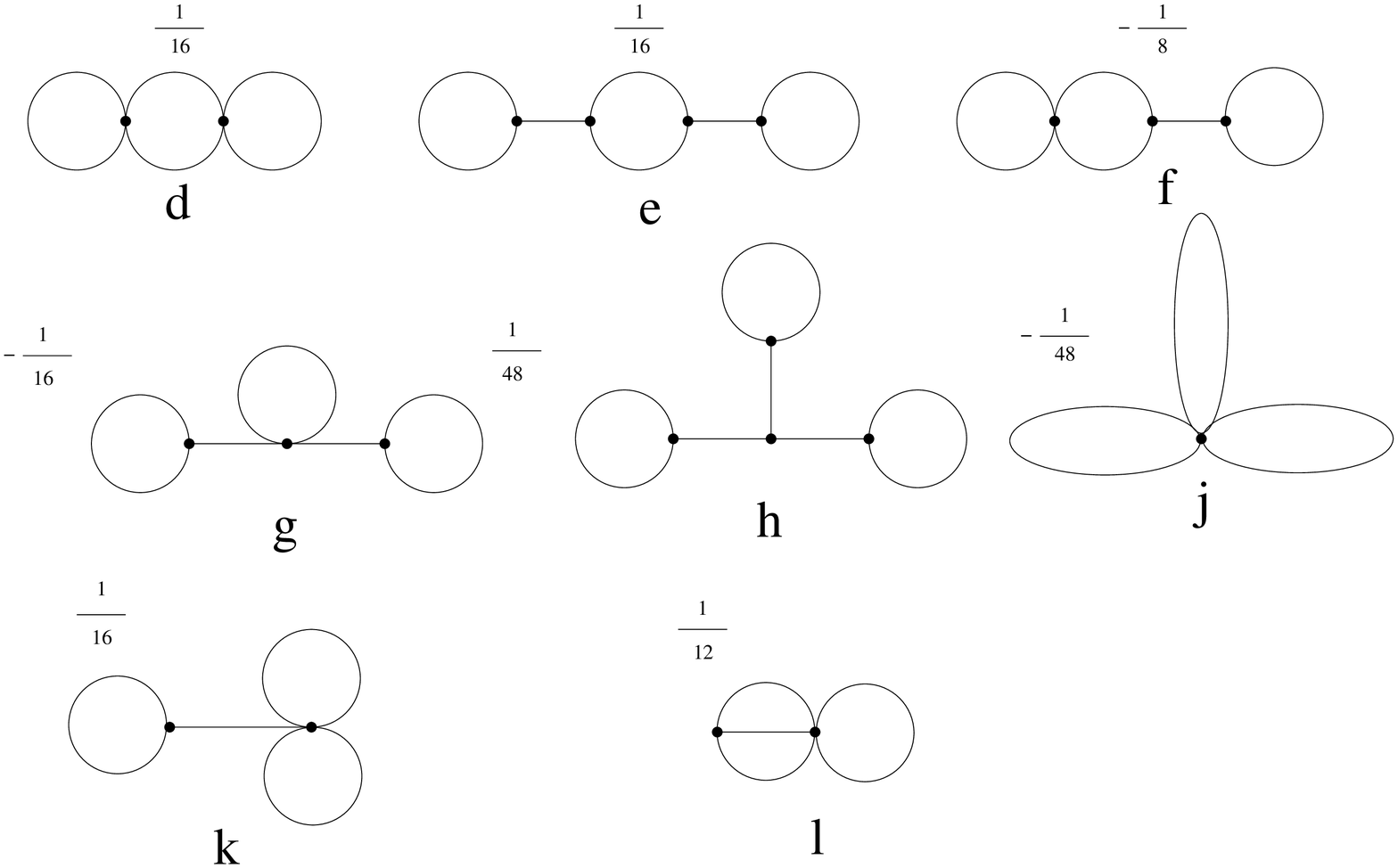}
\caption{Diagrams contributing to the coefficient $B_{2}$. Triple, quartic, quintic and sextic vertices $V_3, V_4, V_5, V_6$, all are marked by (filled) bullets. The signs of contributions and symmetry factors are indicated.}
\label{F2}
\end{center}
\end{figure}

\begin{figure}[htp]
\begin{center}
\includegraphics[width=4.5in,angle=0]{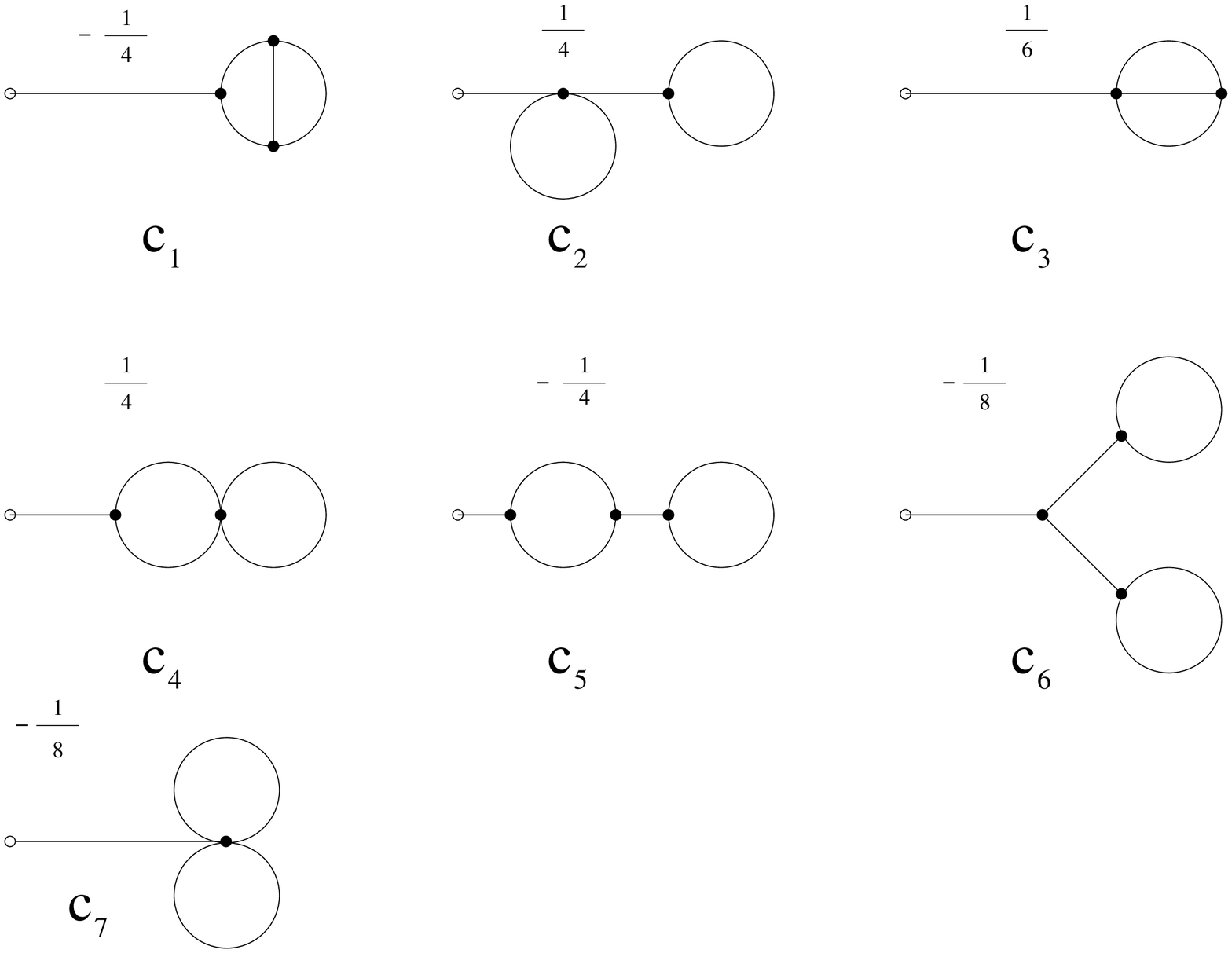}
\caption{Tadpole diagrams contributing to the coefficient $B_{2}$. They come from the Jacobian of the zero mode and have no analogs in the anharmonic oscillator problem. Tadpole vertex $V_{tad}$ (Jacobian source) is marked by (unfilled) open bullet.
The signs of contributions and symmetry factors are indicated.}
\label{F3}
\end{center}
\end{figure}

The three-loop correction $B_2$ (\ref{delE0}) we are interested in is given by the
sum of twenty-two 3-loop Feynman diagrams, which we group as follows
(see Figs. \ref{F2} - \ref{F3})
\begin{equation}
\begin{aligned}
 B_{2}  \ = &\  a_1+b_{11}+b_{12}+b_{21}+b_{22}+b_{23}+b_{24}
\\ &  +d+e+f+g+h+j+k+l +c_1+c_2+c_3+c_4+c_5+c_6+c_7 + B_{2loop} \ ,
\label{B2}
\end{aligned}
\end{equation}
\bigskip
complementing by a contribution from two-loop Feynman diagrams, see Fig. \ref{F1},
\[
B_{2loop}\ =\ \frac{1}{2}{(a+b_1+b_2)}^2 +(a+b_1+b_2)\,c\ =\ \frac{341}{1152}\ \ ,
\]
(see (\ref{B1})).

The rules of constructing the integrals for each diagram should be clear from an example:
the explicit expression for the Feynman integral $b_{23}$ in Fig. \ref{F2}, which is
\begin{equation}
b_{23} =  32768\,\int_{-1}^{1}dx\int_{-1}^{1}dy\int_{-1}^{1}dz\int_{-1}^{1}dw\,J(x,y,z,w)\,
\bigg(x\,y\,z\,w\,G_{xx}G_{xy}G_{yz}G_{yw}G_{zw}^2 \bigg) \ ,
\end{equation}

while for $c_{4}$ in Fig. \ref{F3} it takes the form
\begin{equation}
c_{4} =  256\,\int_{-1}^{1}dx\int_{-1}^{1}dy\int_{-1}^{1}dz
\frac{x\,y\,(1-6z^2+z^4)}{{(1+x^2)}^2{(1+y^2)}^2{(1+z^2)}^2(1-z^2)}\,G_{xy}\,G_{yz}^2\,G_{zz}  \ ,
\end{equation}
here we introduced notations $G_{xy}\equiv G(x,y),\,G^0_{xy}\equiv G^0(x,y)$  and
$J=\frac{1}{{(1+x^2)}^2}\frac{1}{{(1+y^2)}^2}\frac{1}{{(1+z^2)}^2}\frac{1}{{(1+w^2)}^2}$. Notice that
the $c$\'{}s diagrams come from the Jacobian of the zero mode and have no analogs
in the perturbative vacuum problem.

\hspace{0.4cm} For calculation of the symmetry factors for a given Feynman's diagram we use
the Wick's theorem and contractions, see e.g \cite{Palmer-Carrington}. It can be illustrated
by the next two examples.
For diagram $j$ the three bubbles can be rearranged in $3!$ ways and each propagator, which starts
and ends at the same vertex forming loop, contributes with a factor of two giving a symmetry factor
of $2^3\times 3! = 48$ . For the diagram $c_3$ with no bubbles the three propagators which connect
the same pair of vertices can be rearranged in $3!=6$ ways only.

\section*{Results}

The obtained results are summarized in Table \ref{Tab1}. All diagrams are of the form of one-dimensional, two-dimensional,
three-dimensional and four-dimensional integrals. The five diagrams $b_{11},\, d,\,k,\,l,\,c_7$,
in particular (see Fig. \ref{F2})
\begin{equation}
\begin{aligned}
& b_{11}  \ = \  \frac{16}{3}\int_{-1}^{1}dx\int_{-1}^{1}dy\,\frac{1}{(1-x^2)(1-y^2)}
\bigg(\ \frac{(1-6x^2+x^4)(1-6y^2+y^4)}{{(1+x^2)}^2{(1+y^2)}^2}\,G^4_{xy} -{(G^0_{xy})}^4 \ \bigg)
\\
& d \ \ \ = \  16 \int_{-1}^{1}dx\int_{-1}^{1}dy\,\frac{1}{(1-x^2)(1-y^2)}
\\
&
  \hskip 4cm \bigg( \frac{(1-6x^2+x^4)(1-6y^2+y^4)}
 {{(1+x^2)}^2{(1+y^2)}^2} G_{xx}G^2_{xy}G_{yy}-G^0_{xx}{(G^0_{xy})}^2G^0_{yy} \bigg) \ ,
\end{aligned}
\end{equation}
correspond to two-dimensional integrals and together with diagram $j$, a one-dimensional integral,
are the only ones which we are able to calculate analytically
\begin{equation}
\begin{aligned}
\vspace{0.3cm}
& b_{11}  \ = \ -\frac{189199}{756000} + \frac{1}{900}\bigg(178\,\zeta(2)  -
204\,\zeta(3)  + 27\,\zeta(4)  \bigg)\  \equiv b_{11}^{rat} + b_{11}^{irrat} \\
& d \  \  \ = \   -\frac{73931}{47250} + \frac{289}{900}\zeta(2)\ \equiv d^{rat} + d^{irrat}\ , \\
& j \quad    = \    \frac{184}{315}  \ ,  \qquad k \    \ = \   -\frac{379}{630} \ ,   \qquad
 l \   \ = \    -\frac{244}{945} \ ,   \qquad    c_7   \ = \    \frac{16}{45}\ ,
\label{b11}
\end{aligned}
\end{equation}
here $\zeta(n)$ denotes the Riemann zeta function of argument $n$ (see \cite{WW:1927}).
Diagrams $b_{11},\, d,$ contain a rational and an irrational contribution such that
\[
\frac{b_{11}^{irrat}}{b_{11}^{rat}}\approx -0.341 \qquad , \qquad \frac{d^{irrat}}{d^{rat}}\approx
-0.338 \ .
\]
It shows that for diagrams $b_{11}$ and $d$ the rational contribution is three times larger than
the irrational part.
In the case of the DW potential the situation is opposite, the irrational part is dominant
(see \cite{DWP}).
Other diagrams, see Table I, were evaluated numerically with an absolute accuracy $\sim10^{-9}$.
Surprisingly, almost all of them (20 diagrams out of 22 ones in total) are of order
$\geq10^{-1}$ as for $B_2$ itself with two of them (diagrams $b_{12},\,b_{21}$)
which are of order $10^{-2}$.

Dunne-{\"U}nsal (see \cite{V. Dunne} and references therein) reports a value of
\begin{equation}
B_2^{Dunne-{U}nsal}=-\frac{59}{128} \ = \  -0.4609375   \ ,
\label{B2Z}
\end{equation}
while present calculation shows that
\begin{equation}
B_2^{present} \approx \ -  0.460937498\ ,
\label{B2O}
\end{equation}
which is in agreement, up to the precision employed in the numerical integration.

Similarly to the two-loop correction $B_1$ the coefficient $B_2$ is negative.
For not-so-large barriers ($S_0\sim1$), the two-loop and three-loop corrections are
of the same order of magnitude.

The dominant contribution comes from the sum of the two-vertex diagrams $d,b_{11},k,l,c_7$
while the four-vertex diagrams $b_{12},b_{21},b_{23},e,h,c_1,c_5,c_6$ provides minor contribution,
the absolute value of their sum represents less than $0.2\%$ of the total correction $B_2$. Interestingly
for both two and three loop cases the largest contribution comes from the 'ears'-like diagrams $a$ and $d$, respectively, $\frac{a}{B_1}\approx 1.01$ and $\frac{d}{B_2}\approx 2.25\ $.

We already noted that individual three-loop diagrams contain irrational numbers.
Since the Dunne-\"Unsal's result is a rational number,  then there must be a cancelation of these
irrational contributions in the sum  (\ref{B2}). From (\ref{b11}) we note that the term $(b_{11}^{irrat}+d^{irrat})$ gives a contribution of order one to the mentioned sum (\ref{B2}),
and therefore the coincidence in the order of $10^{-9}$ between present result (\ref{B2O}) and
one of Dunne-\"Unsal (\ref{B2Z}) is an indication that such a cancelation occurs. Now, we evaluate the coefficients $A_1$, $A_2$ in (\ref{E0}) using Feynman diagrams
(see \cite{Bender}). In order to do it let us consider the Sine-Gordon potential $V \, =\, \frac{1}{g^2}[\,  1- \cos(g\,x) \,  ]  $ and calculate the transition amplitude $\langle x = 0 | e^{-H\, \tau}  | x = 0  \rangle$.
All involved Feynman integrals can be evaluated analytically. In the limit
$\tau \rightarrow \infty$ the coefficients of order $S_0^{-1}$ and $S_0^{-2}$ in front of $\tau$ gives us the
value of $A_1$ and $A_2$, respectively. As it was mentioned above the $c$\'{}s diagrams do not exist in this case. The Feynman integral $a$ in Fig. \ref{F1} give us the value of $A_1$, explicitly it is equal to $$A_1=-2\ .$$ The diagrams $b_{11},d$ and $j$ in Fig. \ref{F2} determine $A_2$, $b_{11}=-\frac{4}{3}$ $d=-8$ and $j=\frac{16}{3}$. Then
$$\ A_2=-4\ ,$$ which is in agreement
with the results obtained in standard multiplicative perturbation theory
(see e.g. \cite{Turbiner:1979-84}).
No irrational numbers appear in the evaluation of $A_1$ and $A_2$.
\begin{table}[th]
\begin{center}
\setlength{\tabcolsep}{20.0pt}
\begin{tabular}{|c  ||  c  | }
\hline
Feynman                     &    Contribution to
\\[1pt]
diagram                     &     $B_2$
\\[4pt]
\hline
$a_{1}$           \quad     &  \quad $ 0.076497376  $
\\[3pt]
\hline
$b_{12}$          \quad     &  \quad $ 0.012018080   $
\\[3pt]
\hline
$b_{21}$          \quad     &  \quad $0.021795546   $
\\[3pt]
\hline
$b_{22}$          \quad     &  \quad $0.126625453  $
\\[3pt]
\hline
$b_{23}$          \quad     &  \quad $ 0.089343699 $
\\[3pt]
\hline
$b_{24}$          \quad     &  \quad $ 0.136767200 $
\\[3pt]
\hline
$e$               \quad     &  \quad $ 0.095677222  $
\\[3pt]
\hline
$f$               \quad     &  \quad $0.316847853  $
\\[3pt]
\hline
$g$               \quad     &  \quad $ 0.308988525  $
\\[3pt]
\hline
$h$               \quad    &   \quad $ 0.061784864  $
\\[3pt]
\hline
$c_1$             \quad    &    $-0.053695603  $
\\[3pt]
\hline
$c_2$             \quad     &   $-0.347282176 $
\\[3pt]
\hline
$c_3$             \quad     &   $-0.071612992  $
\\[3pt]
\hline
$c_4$             \quad     &   $-0.183024909 $
\\[3pt]
\hline
$c_5$             \quad     &   $-0.114906259 $
\\[3pt]
\hline
$c_6$             \quad     &   $-0.111263501 $
\\[3pt]
\hline
$I_{2D}$           \quad    &   $\quad  -1.70563$
\\[3pt]
\hline
$I_{3D}$           \quad    &   $\qquad 0.36381$
\\[3pt]
\hline
$I_{4D}$           \quad    &   $\qquad 0.00075$
\\[3pt]
\hline
\end{tabular}
\caption{Contribution of each diagram in Fig. \ref{F2} - \ref{F3} for the three-loop correction $B_2$ with symmetry factor included. We write $B_{2} = (B_{2loop} + I_{1D}+I_{2D} +I_{3D} +I_{4D}) $ where $j=I_{1D}$ and $I_{2D},I_{3D},I_{4D}$
denote the sum of two-dimensional, three-dimensional and four-dimensional integrals, respectively.
The term $B_{2loop}=341/1152\approx 0.296 $ (see text).}
\label{Tab1}
\end{center}
\end{table}

\section*{Conclusions and Discussion}
In conclusion, we have calculated the tunneling amplitude (level splitting related to the
instanton density) up to three-loops using Feynman diagrams for quantum perturbations
on top of the instanton. Summing all of these contributions we obtain the third coefficients
$B_2$ (defined in (\ref{delE0})). The result -- to the numerical accuracy we kept -- is found
to be in good agreement with the resurgent relation between perturbative and instanton series
suggested by J.~Zinn-Justin and collaborators, and Dunne-\"Unsal (for modern reference see \cite{V. Dunne}).

\hspace{0.4cm}
Let us remind again, that this paper is methodical in nature, and its task was to
develop tools to calculate tunneling phenomena in multidimensional QM or QFT context, in which
any results stemming from the Schr\"odinger equation are not available. We use a quantum mechanical
example as a test of the tools we use: but the tools themselves are expected to work in much wider context.

\hspace{0.4cm}
When we started these works (see \cite{DWP}) we, naively, expected to see some correspondence between vacuum and instanton series on the level of individual Feynman diagrams. However, no such trend has been detected so far. Furthermore, ``new"  diagrams originating from the instanton zero mode Jacobian, surprisingly, provide the significant contribution $\sim 50\%$ to the two-loop correction $B_1$ (one diagram out of four, see Fig.1) and $\sim 114\%$ to the three-loop correction $B_2$ (seven diagrams out of 22, see Figs.2-3), see Table I, both for 2-loop and 3-loop contributions, both for the double well \cite{DWP} and SG problems. In the double well case \cite{DWP} the ``new" individual diagrams $c$ and $c_5$ give the dominant contribution ($\sim 83\%$ and $\sim 127\%$) to the overall loop coefficients $B_1$ and $B_2$ out of 4 and out of 18 diagrams, respectively, while in the SG case they give significant contributions $\sim 48\%$ and $\sim 25\%$ out of 4 and out of 22 diagrams, respectively. However, the corresponding $c_5$-like four-loop diagram in the SG case represents the $\sim 4\%$ of the four-loop correction $B_3$ only. We calculated $c_2$-like three-, four-, five-loop diagrams (see Fig. 3 for 3-loop case and {\it Note Added} as of Nov.1, 2015, see below): This single tadpole diagram gives $\sim 75\%$, $\sim 50\%$ and $\sim 30\%$ contribution for three-, four-, five-loop  cases, respectively. It is quite amusing that in three-loop case the sum of $c_2$ and $C_5$ diagrams gives $\sim 100\%$. It implies that the sum of remaining 20 diagrams is almost zero!

\hspace{0.4cm}
Another observation  is that the final three-loops answer
has a rational value. However, unlike the evaluation of the two-loop coefficient $B_1$ where
all Feynman diagrams turned out to be rational numbers, in our case of  $B_2$ at least two diagrams
contain irrational parts. What is the origin of these terms and how they cancel out among themselves
are questions left unanswered above, since several diagrams had resisted our efforts to get
the analytic answer, so that we used numerical multidimensional integration methods, in particular,
a dynamical partitioning \cite{PR:2006}. Perhaps, this can still be improved.

\hspace{0.4cm}
Similar calculations for scalar and eventually gauge theories would be certainly
possible and are of obvious interest. The diagrams are the same, and the basic element
remains explicit Green functions. (In the case of gauge theories those should be
orthogonal to all --including gauge-- zero modes.)

\begin{acknowledgments}
MAER is grateful to J.C.~L\'opez Vieyra for assistance with computer calculations and
for the kind permission to use the code for dynamical partitioning in multidimensional
integration.
This work was supported in part by CONACYT grant {\bf 166189}~(Mexico) for MAER and AVT,
and also by DGAPA grant {\bf IN108815-3} (Mexico) for AVT.
The work of ES is supported in part by the U.S. Department of Energy under Contract
No. {\bf DE-FG-88ER40388}.

\bigskip

\textit{Note added in proof (July, 2015)}:

Note that the normalization of the Green functions $G^0,G$ in this paper is
different from \cite{DWP}.
After the paper was submitted we obtained additional
results. Using Dunne-\"{U}nsal procedure we calculated the
$B_3,B_4$ coefficients in the expansion (\ref{delE0})
\[
B_3 = -897/1024
\]
\[
B_4 = -75005/32768
\]
which again turned to be rational. We evaluated contributions of the
$c,c_5$-like diagrams, with maximal number of integrations,
to the four and five loop coefficients. Those diagrams still contribute
significantly to the total answer, although the number of
diagrams grows dramatically with order. 
Those are $\sim 48\%, \sim 25\%, \sim 4\%, \sim 0.4\%$
of total two-, three-, four-, five-loop $B_1,B_2,B_3,B_4$ coefficients, respectively.
Unlike the case of the double-well potential \cite{DWP} the absolute values
of these diagrams tend to decrease by factor $\sim 3$ with each order. 

\bigskip

\textit{Note added (November, 2015)}:

We evaluated contributions of the $c_2$-like diagrams (see Fig.3), with maximal number of four-point vertices $V_4$, to the four and five loop coefficients.
Those are $\sim 75\%, \sim 50\%, \sim 30\%$ of total three-, four-, five-loop $B_2,B_3,B_4$ coefficients, respectively. These anomalously large contributions need to be explained. 

\end{acknowledgments}

\newpage

\clearpage


\begin{thebibliography}{99}

\bibitem{DWP}
         M.A. Escobar-Ruiz, E. Shuryak, A.V. Turbiner,
         {\em Three-loop Correction to the Instanton Density. I. The Quartic Double Well Potential},\\
         { arXiv:1501.03993v3  } (updated, August 2015)\\
         {\em Phys. Rev. D \bf 92}, 025046 (2015)

\bibitem{Scott}
           A. C. Scott, F. Y. F. Chu, and D. W. McLaughlin,
           {\em Proc. IEEE \bf 61}, 1443 (1973)

\bibitem{ColemanI}
           S.~Coleman,
           {\em Phys. Rev. D \bf 11}, 2088-2097 (1975)

\bibitem{Liang}
           J.-Q. Liang and H. J. W. Miiller-Kirsten,
           {\em Phys. Rev. D \bf 51}, 718-725 (1995)

\bibitem{J. Zinn-Justin}
         J.~Zinn-Justin and U.D.~Jentschura,
         {\em Annals Physics \bf 313}, 269-325 (2004)\\
         quant-ph/0501137 (updated, February 2005)

\bibitem{V. Dunne}
         G.~V.~Dunne and M.~\"{U}nsal,
         {\em Phys. Rev. D \bf 89}, 105009 (2014)

\bibitem{Stone}
         M. Lowe and M. Stone,
         {\em Nucl. Phys. B \bf 136}, 177-188 (1978)

\bibitem{Palmer-Carrington}
         C.D.~Palmer, M.E.~Carrington,
         {\em Can. J. Phys. \bf 80}, 847-854 (2002);\\
         P.~V.~Dong et al.
         {\em Theor. Math. Phys. \bf 165}, 1500-1511 (2010)\\
         {arXiv:0907.0859}

\bibitem{Bender}
         C.~M.~Bender and T.~T.~Wu,
         {\em Phys. Rev. \bf 184}, 1231-1260 (1969)

\bibitem{1-E. Shuryak}
         A.~A.~Aleinikov and E.~Shuryak,
         {\em Sov. J. Nucl. Phys. \bf 46}, 76 (1987)

\bibitem{WW:1927}
        E.T.~Whittaker and G.N.~Watson,
        \textit{A Course in Modern Analysis}\\
        4th edition, Cambridge University Press, 1927

\bibitem{Turbiner:1979-84}
         A.~V.~Turbiner,
         {\em JETP Lett. \bf 30}, 352-355 (1979)
           (English Translation)\\
         {\it Soviet\ Phys.\ -- \ Pisma\ ZhETF \bf 30}, 379-383 (1979)

\bibitem{PR:2006}
         A.~V.~Turbiner and J.C.~L\'opez Vieyra,
        {\it Physics Reports \bf 424}, 309-396 (2006)

\end{thebibliography}
\end{document}